\documentclass[aps,prb,twocolumn,showpacs,superscriptaddress,amssymb]{revtex4}
\usepackage{graphicx}
\usepackage{amsmath}
\usepackage[usenames]{color}
\usepackage{hyperref}
\usepackage{natbib}
\usepackage{bm}
\usepackage{threeparttable}

\begin{document}

\title{Aperiodic dynamical decoupling sequences in presence of pulse errors}

\author{Zhi-Hui Wang, V. V. Dobrovitski}
\affiliation{Ames Laboratory, Iowa State University, Ames, IA, 50011}

\date{\today}
\begin{abstract}
Dynamical decoupling (DD) is a promising tool for preserving
the quantum states of qubits. However, small imperfections in the control pulses can
seriously affect the fidelity of decoupling, and qualitatively change the evolution
of the controlled system at long times. Using both analytical and numerical
tools, we theoretically investigate the effect of the pulse errors accumulation
for two aperiodic DD sequences,
the Uhrig's DD (UDD) protocol [G. S. Uhrig, Phys. Rev. Lett. {\bf 98}, 100504 (2007)],
and the Quadratic DD (QDD) protocol
[J. R. West, B. H. Fong and D. A. Lidar, Phys. Rev. Lett {\bf 104}, 130501 (2010)].
We consider the implementation of these sequences using the electron spins of
phosphorus donors in silicon, where DD sequences are applied to suppress dephasing
of the donor spins.
The dependence of the decoupling fidelity on different initial states
of the spins is the focus of our study. We investigate in detail the initial drop
in the DD fidelity, and its long-term saturation.
We also demonstrate that by applying the control pulses along different
directions, the performance of QDD protocols can be noticeably improved,
and explain the reason of such an improvement.
Our results can be useful for future implementations of the aperiodic decoupling
protocols, and for better understanding of the impact of errors on quantum control
of spins.
\end{abstract}


\maketitle

\section{Introduction}

Mitigating the effect of decoherence is an important problem in the
emerging area of quantum information processing (QIP)\cite{NielsenChuang} and other quantum-based technologies
\cite{qrepeater,qmetrology1,qmetrology2,qmetrology3,qmagnetNV1,qmagnetNV2,qmagnetNV3,qmagnetNV4}.
Dynamical decoupling (DD) is a promising tool for this task \cite{Viola98,Viola99}.
Originating from the ideas that underlie the spin echo effect in nuclear magnetic resonance (NMR) \cite{Hahn50,Slichter},
DD employs a specially designed sequence of control pulses
applied to the qubits (or the central spins) in order to negate the coupling of the central spins to their
environment.
A variety of DD protocols have been developed, analyzed, and employed for a few decades
in the area of high-resolution NMR \cite{Haeberlen,GersteinDybowski}.
It has been suggested later that the decoupling property of these
protocols can be used for more general purpose of suppressing decoherence and achieving
high-fidelity quantum operations, thus giving rise to extensive
theoretical \cite{Viola99,Khodjasteh05,RBLiu07,Zhang07,Uhrig07,Cywinski08,Cywinski09,Uhrig08,Zhang08,Zhang08a,West10,Khodjasteh10,Pasini10}
and experimental \cite{Beavan05,Morton08,Ladd08,Uys09,Biercuk09,Du09,Dobrovitski10,Naydenov10,Suter10,deLange10,Ryan10,Bluhm10,Tyryshkin10}
investigation of various novel aspects of DD in the QIP context.
Since DD requires no feedback and no ancilla resources, and
is applicable to a wide range of systems, it presents a promising
tool for lowering the number of quantum errors beyond the error
correction threshold \cite{NielsenChuang}.

Performance analysis of the decoupling protocols is often
based on the Magnus expansion (ME) \cite{Haeberlen}, which is
an asymptotic cumulant expansion of the
evolution operator of the system, with the
characteristic inter-pulse delay (or duration of a single DD cycle, for
periodic sequences) playing the role of the small expansion parameter.
By designing the protocols which nullify more and more expansion terms,
one expects that in favorable experimental situations the decoupling
fidelity increases.
An example of the first-order decoupling protocol is a very basic
periodic dynamical decoupling (PDD) sequence \cite{Santos05,Viola99},
which consists of repeating the basic cycle
$d-\pi-d-\pi$, where $d$ denotes the free evolution of the system, and
$\pi$ denotes a 180$^\circ$ spin rotation around e.g.\ $x$-axis (all analysis
here and below is performed in the rotating frame, i.e.\ in the coordinate
frame rotating around the $z$-axis with the Larmor precession frequency \cite{Slichter}).
This sequence works for the Ising-type coupling, for instance, when a central spin
is coupled to the decohering bath via
the term $S^z B$, where $S^z$ is the $z$-component of the central spin
and $B$ is the bath operator. Such terms lead to pure dephasing of the
central spins in the $x$-$y$ plane, i.e.\ only transverse components of
the central spins are affected by the bath. In order to eliminate more general type
of coupling, involving decoherence of all components of the central spin, a two-axis control
is needed, when the 180$^\circ$ rotation axes alternate between $x$ and $y$ directions \cite{Viola99},
so the basic period of the PDD sequence has a form $d-\pi_X-d-\pi_Y-d-\pi_X-d-\pi_Y$,
with $\pi_X$ and $\pi_Y$ denoting the $\pi$ rotations around $x$ and $y$, respectively.
To increase the decoupling order, eliminating the second-order ME terms as well,
we need to use more advanced sequences than PDD. One possibility is to use symmetrized
sequences (SDD) \cite{Haeberlen,Viola99,Santos05} having a twice longer cycle,
consisting of the PDD cycle followed by its mirror image. Another approach
to achieve higher-order decoupling is to use concatenated DD (CDD) protocol, which
removes the system-bath coupling terms in the ME up to the $n$-th order
by nesting the PDD sequence recursively to itself $n$ times.\cite{Khodjasteh05,Khodjasteh07,Zhang07,Zhang08}
The number of pulses in CDD increases exponentially with
the concatenation level, approximately as $4^n$.
Within CDD approach, the total decoupling time can be increased either by
increasing the concatenation order, or by periodically repeating the
fixed-level CDD protocol.

Another way to improve the decoupling fidelity is to optimize positions of the
pulses, thus creating a family of aperiodic sequences where the inter-pulse delays
are not necessarily commensurate.
In Ref.~\onlinecite{Uhrig07} G.~S.~Uhrig proposed a sequence (later abbreviated as UDD),
where optimization of the pulse positions provides $n$-th order decoupling
using $n$ pulses \cite{Lee08,WYang08}.
UDD has been implemented and tested experimentally in different systems \cite{Biercuk09,Uys09,Du09,deLange10,Ryan10,Suter10}.
Theoretical studies have shown that UDD is optimal when the
noise spectrum of the bath has a sharp high-frequency cutoff \cite{WYang08,Biercuk09,Du09},
while for the spectra with soft cutoffs the protocols with periodic structure,
such as the Carr-Purcell-Meiboom-Gill \cite{Slichter} (CPMG) sequence,
perform better \cite{Pasini10,Cywinski08,deLange10,Ajoy10}.
While UDD employs a single-axis control, and therefore suppresses only pure dephasing,
several UDD-based aperiodic protocols were developed to suppress general decoherence
by incorporating double-axis controls,
including the concatenated version of UDD (CUDD) \cite{UhrigCUDD09},
and quadratic DD (QDD) where two UDD sequences based on spin rotations
about perpendicular axes are nested into each other \cite{West10}.
In CUDD, the number of pulses also grows exponentially
but slower than CDD, and in QDD the pulse number required
to suppress $n$-order decoherence is of order of $O(n^2)$.


While DD protocols are usually designed based on the assumption of ideal pulses,
in reality the accumulation of small imperfections
of the pulses can severely affect the decoupling fidelity and even qualitatively
change the evolution of the central spins.
The influence of the pulse errors on the DD protocols has been
studied from early days of NMR \cite{GersteinDybowski,Vaughan72} within the ME (or ME-like) settings,
by considering the pulse errors as extra terms in the Hamiltonian. However,
such an approach is not always satisfactory. First, the convergence conditions of ME are not
always satisfied in the experiments, although the DD may still efficiently suppresses
decoherence. Second, due to its asymptotic nature, the ME can miss the possibility
of disastrous pulse error accumulation
at long times, when the number of pulses becomes very large.
Very detailed studies of the errors introduced by the finite pulse
width have been performed for CDD \cite{Khodjasteh07} and UDD \cite{UhrigPasiniNJP10,Biercuk09}.
The experimental study of the systematic errors in the rotation axis and angle
has been done for a number of periodic protocols and for UDD \cite{Suter10,Biercuk09PRA},
and both error types (finite-width and systematic) have been discussed for
dynamically corrected gates \cite{KhodjastehDCG09}.
The theoretical-experimental investigation of the long-term accumulation of the systematic
pulse errors has been done for several periodic-based protocols,
PDD, SDD, and CDD \cite{Tyryshkin10,Wang10}.

In this work, we present a theoretical analysis of the aperiodic protocols,
UDD and QDD, focusing on the long-term accumulation of the systematic pulse
errors. These errors can be much more devastating than the imperfections associated
with the finite width of the pulses.
We use numerical modeling to analyze the situation when the aperiodic
protocols with imperfect pulses are used for decoupling of the electron spins of
the phosphorus donors in silicon. This system may present a promising platform
for scalable solid-state quantum information processing, and is a good testbed
for studying a number of fundamental issues in quantum dynamics and quantum
control of solid-state spins \cite{Kane98,Tyryshkin06,Stegner06,Hollenberg10}.
The Si:P system is well characterized \cite{Feher59,Honig56,Tyryshkin03,Fletcher},
DD of the electron spins of P donors is efficiently implemented via pulsed ESR technique,
and the realistic pulse error parameters have been estimated earlier \cite{Tyryshkin10,Wang10}.
Comparing the results of our study below with the previous works on periodic
protocols in the same system, we find that the aperiodic sequences provide
reasonable decoupling fidelity, but they are more demanding to the pulse error
magnitude than e.g.\ CDD.
We also demonstrate that, by choosing a right set of two-axis controls,
the effect of the pulse errors can be significantly reduced, and explain the
reason behind this improvement. It is interesting
that the right choice for QDD protocol, employing rotations around the $z$ and $y$ axes,
is a bad choice for CDD and PDD, where the rotations around $x$ and $y$ should be
used \cite{Tyryshkin10,Wang10}. It demonstrates again that the analysis of
the pulse error accumulation is, in general, protocol-specific.

The rest of the paper is organized as follows.
In Sec.~\ref{sec:system}, we describe the Si:P system and present a model for
the pulse errors.
In Sec.~\ref{sec:results}, we present the analytical and numerical results for UDD and QDD.
Conclusions are given in Sec.~\ref{sec:conclusions}.

\section{Description of the system}
\label{sec:system}
\subsection{The Phosphorus Doped Silicon System}
The electron spins of P donors in silicon have long relaxation and coherence times \cite{Honig56,Feher59,Tyryshkin03},
and have been used earlier to explore fundamental aspects of quantum spin
dynamics and DD \cite{deSousa03,Witzel05,Tyryshkin03,Tyryshkin06,Morton08,Tyryshkin10,Wang10,Witzel10}.
In this study, we consider the same experimental situation as described in Ref.~\onlinecite{Tyryshkin10}.
The sample is an isotopically purified bulk silicon
with $^{29}$Si concentration $\sim800~$ppm, and very low doping density of phosphorus, so that
the coupling between different P donor spins can be neglected at the relevant timescales.
A large static (quantizing) magnetic field is applied along the $z$-axis to the system,
and the temperature is sufficiently low ($8~$K in the experiments reported in
Ref.~\onlinecite{Tyryshkin10}). The relaxation time $T_1$ of the P electron spin is
very long, so the longitudinal relaxation is neglected here. Similarly, the
coherence time $T_2$, defined as the spin echo decay time, is long (several milliseconds, limited by instantaneous
diffusion) and will also be neglected below \cite{Tyryshkin03,Tyryshkin06,Tyryshkin10}.

Thus, we consider only pure dephasing of the P electron spins. One important dephasing channel
is the Ising-type hyperfine coupling to the $^{29}$Si nuclear spins \cite{Honig56,Feher59,Tyryshkin03,deSousa03,Witzel05}.
Due to rather large localization radius of the P donor electron in silicon, the
P electron spin couples to a large number (bath) of $^{29}$Si nuclear spins.
For an isotopically purified sample considered here, the interactions between different
$^{29}$Si nuclear spins are small and
can be neglected, so we are dealing with the static nuclear spin bath.
Moreover, since the P donors are well separated,
the $^{29}$Si nuclei which are appreciably coupled to
one P electron spin interact very weakly with other donor centers,
so each donor can be considered as a separate central spin with its own
static spin bath.
Another important contribution to dephasing is the inhomogeneous
broadening due to quasistatic fluctuations of the quantizing magnetic field.
Therefore, in the frame rotating
with the frequency at the center of the ESR line of the P electrons,
the Hamiltonian for a P electron spin coupled to the bath is
\begin{equation}
{\mathcal H}=B S^z
\end{equation}
where $S^z$ is the $z$-component of the electron spin operator,
and $B$ is the total random static offset field created by the
bath of $^{29}$Si nuclei and the fluctuations of the external field \cite{Slichter,Abragam}.
Experiments show that the ESR lineshape is Gaussian with the width of 50~mG,
so that the distribution function of $B$ is
\begin{equation}
P(B)=\frac{1}{\sqrt{2\pi b^2}}\exp{\left[-\frac{B^2}{2b^2}\right]}
\end{equation}
with $b=50$~mG. Note however, that our results are not very sensitive to the
specific shape of the distribution.
The hyperfine coupling of the P electron to the donor's own $^{31}$P nuclear spin ($I=1/2$)
is large \cite{Feher59,Fletcher}, of order of 100~MHz, so that the two hyperfine lines
are well separated, and we assume that only one line is excited. Then the
on-site hyperfine coupling to $^{31}$P nucleus just shifts the ESR frequency, and
we neglect this interaction.

As is standard in NMR and ESR, the experimental temperature is much larger than
the Zeeman energy of the electron spins in the quantizing field. Thus,
we describe the state of a P spin within the high-temperature approximation,
as is customarily done in NMR/ESR theory \cite{Slichter}, using the
density matrix $\rho\propto {\mathbf 1}+\alpha S^z$, where $\alpha$ is
small.
Since the  $T_1$ time is extremely large, and the identity matrix is not affected by the
unital evolution, we can neglect it, and
consider the electron spin as being in a pure state with $S^z=1/2$ (so-called pseudo-pure state).
Other initial (pseudo)pure states, e.g.\ along the $x$ and $y$ axes, can
be prepared by the $\pi/2$ pulses applied along the $y$ and $-x$ axes, respectively.
For pure initial states performance of the decoupling can be conveniently
quantified by the input-output fidelity,
i.e.\ by the overlap between the initial and the final (reduced) density matrices of the system.
By virtue of the relation $\rho(t)=\frac{\mathbf 1}{2}+\langle S^x\rangle \sigma_x
+\langle S^y\rangle \sigma_y+\langle S^z\rangle \sigma_z$,
the average spin projections $\langle S^x\rangle$, $\langle S^y\rangle$, and $\langle S^z\rangle$
can be equivalently
used as measures of the decoupling performance. Note that the angular brackets
here denote both quantum-mechanical averaging (including trace over the bath
states) and the averaging over the ensemble of different P electron spins.
Below, we use the rescaled fidelity, which for the initial state along the axis $\alpha=x,y,z$
is defined as $F_\alpha = 2\langle S_\alpha\rangle$.


\subsection{The analytical model of pulse errors}


For the experimental situation under consideration the systematic errors
in the pulse parameters (rotation angle and the direction of the rotation
axis) are much larger than the errors associated with the finite pulse
width. Therefore, we treat each $\pi$-pulse as an instantaneous but imperfect spin rotation.
For a nominal $\pi$-pulse about the $x$ axis (the $\pi_X$ pulse)
the unitary operator describing the spin evolution is
\begin{eqnarray}
\label{eq:piX}
U_{\text X}&=&\exp{[-i(\pi + \epsilon_x)({\bf S}\cdot\vec {\bf n})]}\\ \nonumber
\end{eqnarray}
where $\epsilon_x$ is the error in the rotation angle, and
$\vec{\bf n}=(\sqrt{1-n_y^2-n_z^2},n_y,n_z)$ is the unit vector along the rotation axis, where
the small parameters $n_y$ and $n_z$
characterize the deviation of the actual rotation axis from the $x$ axis.
Similarly, the rotation operator for a nominal $\pi_Y$ pulse is
\begin{eqnarray}
\label{eq:piY}
U_{\text Y}&=&\exp{[-i(\pi + \epsilon_y)({\bf S}\cdot\vec {\bf m})]}\;,
\end{eqnarray}
where $\epsilon_y$ is the rotation angle error,
and $\vec{\bf m}=(m_x,\sqrt{1-m_x^2-m_z^2},m_z)$ is the actual rotation axis,
with small parameters $m_x$ and $m_z$.
During the delay between neighboring pulses,
the evolution operator is simply
\begin{equation}
\label{eq:Ud}
U_d(\tau)=\exp[-iB S^z \tau]
\end{equation}
where $\tau$ is the duration of the delay between the pulses (e.g., for
periodic sequences studied in Ref.~\onlinecite{Tyryshkin10} equal to 11~$\mu$s).

To model the systematic errors, we take into account that the resonant ac field $B_p$, which rotates the
spins during pulses, is not homogeneous over the sample. We assume for simplicity that this field varies only along one
spatial axis, denoted as $l$, and the sample is located between $l=+d$ and $l=-d$.
We also assume that the sample is placed optimally with respect to the field,
with the sample center $l=0$ located at the maximum of $B_p$.
Then the spatial dependence of $B_p$ has a form $B_p(l)= \bar B_{p} +\Delta B_p[1-3l^2/d^2]$,
where $\bar B_p$ is the average magnitude of the ac field over the sample
and $\Delta B_p$ characterizes the ac field inhomogeneity. If the pulse width $t_p$ is adjusted
to give $\bar B_p t_p=\pi$, then the spins located at different parts of sample will undergo
rotation by different angles. For $\pi_X$ pulse, the resulting rotation angle errors $\epsilon_x$
will have distribution
\begin{equation}
\label{eq:epsdist}
P(\epsilon_x)=(1/2\epsilon_0)[3(1-\epsilon_x/\epsilon_0)]^{-1/2},
\end{equation}
where $-2\epsilon_0\leq\epsilon\leq \epsilon_0$, and $\epsilon_0$ characterizes
the magnitude of the rotation angle errors.
In a similar way, we assume that the $z$-component of the rotation axis $n_z$
has the same distribution, and its magnitude is characterized
by the parameter $n_0$. We also assume that the pulse errors for
$\pi_{X}$ and $\pi_{Y}$ pulses are
the same, taking $\epsilon_x=\epsilon_y$ and $m_z=n_z$. The pulse
error parameters $m_x$ and $n_y$ have different nature: they characterize the
phase of the pulse, and can be adjusted very precisely, so we assume them to be zero,
following Refs.~\onlinecite{Tyryshkin10,Wang10}. Thus, within this model,
all pulse errors are characterized by only two parameters, $\epsilon_0=0.3$ and
$n_0=-0.12$. These values have been determined in Refs.~\onlinecite{Tyryshkin10,Wang10},
and have reasonable magnitude. In spite of utter simplicity, this model
reproduces (or, at least, mimics) well most features observed in experiments.

\section{Numerical and Analytical Results}
\label{sec:results}
\subsection{Performance of UDD}
\begin{figure}[htbp]
\includegraphics[height=8cm, angle=270]{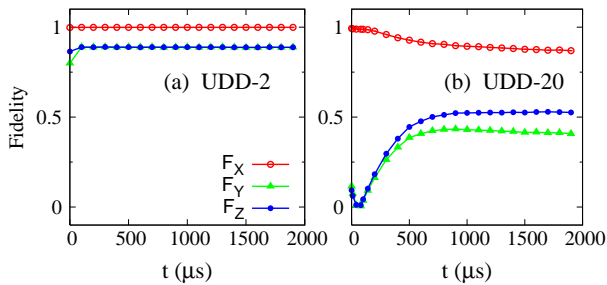}
\caption{\label{fig:UDDeven} (Color online).
Decoupling fidelities of UDD-2 (a) and UDD-20 (b) as functions of the total evolution time
for initial states $S_X$ (red empty circles, red lines), $S_Y$ (green triangles, green lines), and $S_Z$
(blue solid circles, blue lines).
}
\end{figure}

\begin{figure}[htbp]
\includegraphics[height=8cm, angle=270]{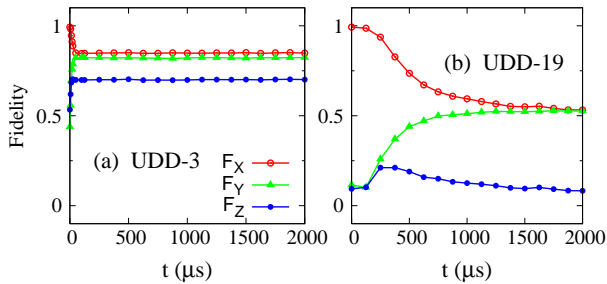}
\caption{\label{fig:UDDodd} (Color online).
Decoupling fidelities of UDD-3 (a) and UDD-19 (b) as functions of the total evolution time
for initial states $S_X$ (red empty circles, red lines), $S_Y$ (green triangles, green lines), and $S_Z$
(blue solid circles, blue lines).
}
\end{figure}

Uhrig's DD protocol, UDD, is based on a single-axis control, i.e., all spin rotations
are performed about the same nominal axis, which we take as $x$.
For UDD of level $\ell$, denoted as UDD-$\ell$, the $\pi_X$ pulses
are applied at times\cite{Uhrig07}
\begin{eqnarray}
\label{eq:UDDtiming}
t_j=t\sin^2\left(\frac{j\pi}{2\ell+2}\right)\;.
\end{eqnarray}
where $t$ is the total evolution time,
and $j=1,2,\ldots,\ell$ for even $\ell$ and $j=1,2,\ldots,\ell+1$ for odd $\ell$.
The simulation results show that UDD of even levels and UDD of odd levels
behave differently.
Fig.~\ref{fig:UDDeven} shows the decoupling fidelities for UDD-$\ell$ as functions of total
evolution time for $\ell=2,~20$,
and Fig.~\ref{fig:UDDodd} shows the results for UDD-$\ell$ with $\ell=3$ and $19$.

We consider fidelities for three different initial states of the central spins,
directed along $x$, $y$, and $z$ axes. They characterize how well the corresponding
spin components are preserved.
For UDD-2, which is the same as a single cycle of CPMG,
the spin component $S_X$ is preserved with fidelity close to 1,
while the fidelities for the other two components are affected by the pulse errors.
For long evolution time, $F_y$ and $F_z$ saturate at values $\sim0.89$.
Strong dependence of the fidelity on the initial state has been noticed
and analyzed before for periodic protocols,
and it is not unexpected that aperiodic sequences demonstrate this feature as
well \cite{Zhang08,Tyryshkin10,Wang10,Suter10}.
For $\ell=20$, $S_X$ is well preserved, demonstrating only a slow decay,
while $F_y$ and $F_z$ have initial values close to zero, and saturate to a value around $0.5$.
Such low values even for $t=0$ are caused purely by a catastrophic accumulation of errors
of 20 $\pi_X$ pulses.

Let us consider the level-2 UDD
as an example in order to examine the long time behavior.
The evolution operator, up to second order in the pulse errors, and using $n_y=0$, is
\begin{equation}
U^{\tt UDD}_{2}=-\big[1-\theta_x^2/2-i\sigma_x\theta_x-i\sigma_z\theta_z\big]
\end{equation}
with $\theta_x=\epsilon_x\cos\frac{Bt}{4}+2n_z\sin\frac{Bt}{4}$
and $\theta_z=\epsilon_xn_z\cos\frac{Bt}{2}+(n^2_z-\frac{\epsilon^2_x}{4})\sin\frac{Bt}{2}$.
Although the qualitative features of the protocol can be analyzed using only the
first-order expression for $U^{\tt UDD}$, quantitative analysis of the fidelity requires the
second-order terms in this evolution operator.
Taking into account that $\epsilon_x$, $n_z$, and $B$ are independent variables,
and that $\langle \epsilon_x \rangle=\langle n_z \rangle=0$,
the expression for $F_y$ becomes
\begin{eqnarray}
F_y &=& 1-2\theta_x^2 \\ \nonumber
&=& 1-2\big[ \langle \epsilon^2_x \rangle \langle \cos^2\frac{Bt}{4}\rangle
+4\langle n^2_z \rangle\langle \sin^2\frac{Bt}{4}\rangle\big].
\end{eqnarray}
Noticing from Eq.~(\ref{eq:epsdist}) that
$\langle \epsilon^2_x \rangle=0.8\epsilon^2_0$, and $\langle n^2_z \rangle=0.8n^2_0$,
we arrive at
%
%
%
%
\begin{equation}
\lim_{t\to\infty} F_y = 1 - 0.8 (\epsilon^2_0+4n^2_0) = 0.88
\end{equation}
This is close to the value 0.89 which we see in the simulations shown in Fig.~\ref{fig:UDDeven} for UDD-2.
The analysis above relies on the fact that $B$ is time-independent, so
in experiments such saturations will be replaced by decays at different rates, depending on the bath dynamics.
Therefore, although our results extend until $t=2000$~$\mu$s, at these times the instantaneous
diffusion and the internal bath dynamics should be taken into account. We neglect these effects
here since they are beyond the scope of our present study, and require a separate focused
research effort.

For UDD of higher level, similar behavior is observed.
In Fig.~\ref{fig:UDDeven}, the fidelities for UDD-20 are generally smaller than for UDD-2,
which reflects more serious accumulation of the pulse errors.
%


An interesting feature, seen in Fig.~\ref{fig:UDDeven}, is that at
longer times, when dephasing becomes noticeable, the fidelities may increase, as if decoherence counteracts
the pulse errors. To understand this, let us consider UDD-2 in the situation when the errors
$n_z$ are absent, and only the rotation angle error $\epsilon_x$ is nonzero. If a spin is prepared
along the $y$ axis, and is subjected to a sequence of imperfect $\pi_X$ pulses,
then its $y$-component rotates around the $x$-axis, and accumulates the rotation angle errors.
But if the random field $B$ moved the spin away from the $y$-axis towards the $x$-axis by the angle $\chi=Bt/4$,
then the spin component along the $x$ axis is not affected by the rotation angle errors, and
the spin is refocused with better precision.
Correspondingly, the fidelity for a fixed value of $B$ demonstrates a modulation proportional
to $-\cos^2{\chi}$, with maximum fidelity achieved at $\chi=\pi/2$.

For UDD of odd level $\ell=2n-1$, the number of pulses is $2n$
since there is a $\pi$ pulse at the end of the whole evolution.
As shown in Fig.~\ref{fig:UDDodd}, for a given UDD-$\ell$,
when the total evolution time is short, $S_X$ is better preserved
than the other two spin components.
At long time, $F_x$ and $F_y$ saturate at similar values,
while $F_z$ saturates at a lower value.
The process of saturation is much slower for $\ell=19$ than for $\ell=3$.

To gain insight into the behavior of odd-level UDD, we consider UDD-3.
The evolution operator for this sequence, up to first order in the pulse errors, is
\begin{equation}
\label{eq:UDDnOdd}
U^{\tt UDD}_3= {\mathbf 1}-i\theta_n\sigma_x-i\eta_n\sigma_y
\end{equation}
with
\begin{eqnarray}
\theta_n&=&\frac{\epsilon_x}{2}\big[ 1+2\cos(B\tau_1)+\cos\big(B(\tau_2-\tau_1)\big)\big]\nonumber\\
&+&n_z\big[2\sin(B\tau_1)+\sin\big(B(\tau_2-\tau_1)\big) \big]
\end{eqnarray}
and
\begin{eqnarray}
\eta_n&=&-n_z\big[ 1-2\cos(B\tau_1)+\cos\big(B(\tau_2-\tau_1)\big) \big]\nonumber\\
&-&\frac{\epsilon_x}{2}\big[ 2\sin(B\tau_1)-\sin\big(B(\tau_2-\tau_1)\big) \big]\;,
\end{eqnarray}
where $\tau_1$ and $\tau_2$ refer to the durations of the first and second inter-pulse delays.
In general, for quantitative calculation of the fidelities, we need also to consider the second-order terms,
but it can be shown that for UDD-3 their contribution is zero.
%
%
At long times, when $\langle \sin(B\tau_1)\rangle=\langle \sin\big(B(\tau_2-\tau_1)\big)\rangle=0$
(and similarly for cosines),
the fidelity $F_y$ acquires a form
\begin{eqnarray}
F_y&=&1-2\langle \theta_n^2 \rangle \nonumber\\
&=&1-2\langle
5n_z^2+\frac{\epsilon_x^2}{4}+(\frac{\epsilon_x^2}{4}-n_z^2)\Big(4\cos^2(B\tau_1) \nonumber\\
& &+\cos^2 \big(B(\tau_2-\tau_1)\big)\Big)\rangle
\end{eqnarray}
and we finally obtain
\begin{equation}
\lim_{t\to\infty}F_y=1-5\langle n_z^2 \rangle-\frac{7\langle\epsilon_x^2\rangle}{4}=0.82,
\end{equation}
which coincides with the result of the numerical simulations.

We now examine the fidelities at $t\to 0$. Since we treat pulses as
instantaneous rotations, this situation should be understood as the limit
where the total evolution time is much smaller than the dephasing time $T_2^*=1/b$,
but much larger than the total width  $2n t_p$ of all pulses.
Decay of fidelity in this case arises only from accumulation of the pulse errors.
The evolution operators for UDD of both even ($\ell=2n$) and odd ($\ell=2n-1$) levels have the same expression
\begin{eqnarray}
\lim_{t\to 0} U^{\tt UDD}_{\ell}&=&(-1)^n \left( 1-in\epsilon_x \sigma_x \right)\;,
\end{eqnarray}
corresponding to a spin rotation about the $x$ axis with the rotation
angle proportional to $n$.
Small differences in the rotation angles for different spins
are scaled by $n$. For large $n$ the spins
become almost uniformly distributed in the $y$-$z$ plane,
hence $F_y$ and $F_z$ are close to zero, while $F_x$ is close to 1, as seen
in Figs.~\ref{fig:UDDeven} and \ref{fig:UDDodd} for $\ell=20$ and $\ell=19$.


\subsection{Performance of QDD}
In order to suppress general decoherence,
DD sequence has to use two-axis control.
An extension of UDD, the quadratic DD (QDD)
has been suggested for this purpose \cite{West10}.
QDD is composed of two nested UDD sequences, with inner and outer sequences
employing the rotations about two different perpendicular axes.
A QDD sequence of order $\ell$, which we denote as QDD-$\ell$, can suppress
decoherence to $\ell$-th order terms with $(\ell+1)^2$ inter-pulse intervals.

\subsubsection{QDD based on $\pi_X$ and $\pi_Y$ pulses}

We first consider the QDD sequence with the $\pi_X$ pulses in the outer hierarchical level,
and $\pi_Y$ pulses in the inner hierarchical level, examining the effects of error accumulation.

Like UDD, QDD of even and odd levels also behave differently.
For $\ell=2n-1$,
the sequence of QDD-$\ell$ is
\begin{equation}
\label{eq:seq_QDDodd}
\text{UDD}_{\ell}^{(\text Y)}(\tau_1)\text -\pi_\text X\text-\text{UDD}_{\ell}^{(\text Y)}(\tau_2)\text -\pi_\text X\cdots\text-\text{UDD}_{\ell}^{(\text Y)}(\tau_{\ell+1})\text -\pi_\text X
\end{equation}
where $\sum_{j=1}^{\ell+1}\tau_j=t$ and
$\text{UDD}_{\ell}^{(\text Y)}(\tau_j)$ denotes UDD-$\ell$ based on $\pi_Y$ pulses with evolution time $\tau_j$.
The division of the total time $t$ into intervals $\tau_j$ follows the same rule as UDD,
i.e., $\tau_j=t_j-t_{j-1}$, with $t_j$ given by Eq.~(\ref{eq:UDDtiming}).
The number of pulses in QDD with $\ell=2n-1$ is $(\ell+1)(\ell+2)$.
%
%
For $\ell=2n$
the sequence QDD-$\ell$ is
\begin{equation}
\label{eq:seq_QDDeven}
\text{UDD}_{\ell}^{(\text Y)}(\tau_1)\text -\pi_\text X\text-\cdots\text-\text{UDD}_{\ell}^{(\text Y)}(\tau_{\ell})\text -\pi_\text X \text- \text{UDD}_{\ell}^{(\text Y)}(\tau_{\ell+1})
\end{equation}
and the number of pulses is $\ell(\ell+2)$.

\begin{figure}[htbp]
\includegraphics[height=8cm, angle=270]{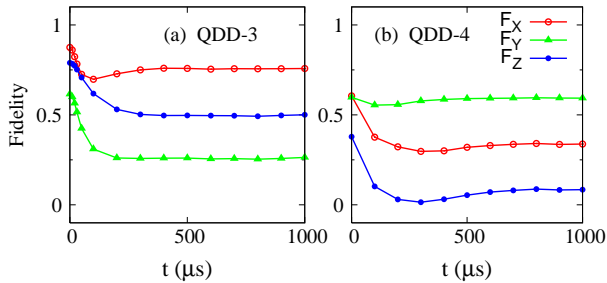}
\caption{\label{fig:QDD} (Color online).
Decoupling fidelities as functions of the total evolution time
for initial states $S_X$ (red empty circles, red lines), $S_Y$ (green triangles, green lines), and $S_Z$
(blue solid circles, blue lines).
(a) QDD-3 (20 pulses). (b) QDD-4 (24 pulses).
}
\end{figure}

Considering the limit $t\to 0$, we have for even and odd levels
\begin{eqnarray}
\label{eq:QDDt0Even}
\lim_{t=0} U^{\tt QDD}_{2n}&=&1-in(\epsilon_x \sigma_x+\epsilon_y\sigma_y)\\
\label{eq:QDDt0Odd}
\lim_{t=0} U^{\tt QDD}_{2n-1}&=&(-1)^n \left( 1-in\epsilon_x \sigma_x \right)
\end{eqnarray}
That is, for short total evolution time,
in a sequence QDD-$(2n)$, the dominant pulse error is $\epsilon_x$,
while in QDD-$(2n-1)$ both $\epsilon_x$ and $\epsilon_y$ contribute in first order.
The difference is due to an extra $\pi_X$ pulse at the end of the odd-order QDD.

%

Simulation results for QDD-3 are shown in Fig.~\ref{fig:QDD} (a).
Fidelities for all three spin components exhibit long-time saturation.
The saturation value for $S_X$ is higher than for the other two components.
However, $F_x$ in QDD-3 exhibits initial decay (even for $t=0$),
instead of being preserved as expected from Eq.~(\ref{eq:QDDt0Odd}).
This is because for the system under consideration, the error parameter
$\epsilon_0=0.3$ is pretty large, and the pulse errors may appreciably contribute in higher orders.
Taking QDD of $\ell=3$ as an example, in the limit $t\to 0$,
the sequence is a fourfold repetition of the unit
\begin{equation}
\label{eq:qdd3example}
\text{($\pi_\text Y$-$\pi_\text Y$-$\pi_\text Y$-$\pi_\text Y$)-$\pi_\text X$}\;.
\end{equation}
Expanding the corresponding evolution operator up to second order
in the pulse errors, and taking into account $m_x=n_y=0$, we find that
\begin{eqnarray}
U=(1-2\epsilon_x^2)-2i\epsilon_x~\sigma_x
+2i\epsilon_x(2\epsilon_y+n_z)\ \sigma_z\;.
\end{eqnarray}
Thus, the rotation axis deviates from the $x$ axis towards the $z$ axis, and
the spin state along the $x$ axis is not exactly parallel to the rotation axis,
resulting in the initial drop of $F_x$.

Fig.~\ref{fig:QDD} (b) shows the simulation results for QDD-4, with 24 pulses.
The saturation values in a descending order are $F_y$, $F_x$ and $F_z$.
The initial values for $F_x$ and $F_y$ are the same,
which can be explained by Eq.~(\ref{eq:QDDt0Even})
taking into consideration that $\epsilon_x=\epsilon_y$ in our model.

As we see, the QDD sequences are rather sensitive to the pulse errors.
While the saturation at long times is seen for all spin components,
the initial fidelity loss is noticeable.

\subsubsection{QDD based on $\pi_Z$ and $\pi_Y$ pulses}

\begin{figure}[htbp]
\includegraphics[height=8cm, angle=270]{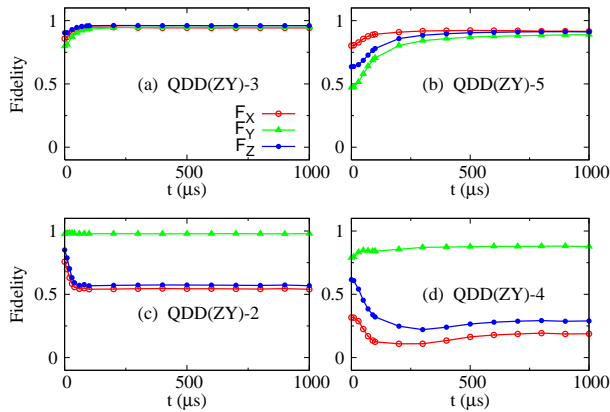}
\caption{\label{fig:ZYQDD} (Color online).
Decoupling fidelities for QDD(ZY) as functions of the total evolution time
for initial states $S_X$ (red empty circles, red lines), $S_Y$ (green triangles, green lines), and $S_Z$
(blue solid circles, blue lines).
(a) level 3 QDD(ZY), (b) level 5 QDD(ZY),
(c) level 2 QDD(ZY), (d) level 4 QDD(ZY).
}
\end{figure}

In order to further understand the effect of pulse errors,
we briefly consider another QDD sequence with $\pi_Z$ pulses in the outer hierarchical level
and $\pi_Y$ pulses in the inner one.
We denote such a sequence as QDD(ZY).
To produce $\pi_Z$ pulses in ESR experiments, the pulses $\pi_X$ and $\pi_Y$
are applied back-to-back \cite{Tyryshkin10}, and our simulations reproduce this situation,
taking into account the $\pi_X$ and $\pi_Y$ pulse errors.
Fig.~\ref{fig:ZYQDD} shows the simulation results for QDD(ZY) of levels 3 and 5 (two upper panels), and 2 and 4
(two lower panels), with 20 and 42, and 8 and 24 pulses, respectively.
While QDD(ZY) of even levels provide no obvious improvement compared to QDD sequences based on $\pi_X$ and $\pi_Y$ pulses,
the QDD(ZY) protocols of odd level preserve all three spin components, with high saturation
values at long times.

To understand this effect, let us inspect the structure of the QDD(ZY) sequence.
As seen from Eq.~(\ref{eq:seq_QDDodd}),
since UDD of odd level has a $\pi$-pulse at the end of the sequence,
after nesting this $\pi_Y$ pulse is followed by a $\pi_Z$ pulse without delay, and
these two adjacent pulses are equivalent to a (imperfect) $\pi_X$ pulse.
The sequence of QDD(ZY) thus has a structure
\begin{eqnarray}
& &
\text{\big($\tau_{1,1}$-$\pi_\text Y$-$\tau_{1,2}$-$\pi_\text Y$-$\tau_{1,3}$-$\pi_\text Y$-$\tau_{1,4}$\big)-$\pi_\text X$-}\nonumber\\
& &
\text{\big($\tau_{2,1}$-$\pi_\text Y$-$\tau_{2,2}$-$\pi_\text Y$-$\tau_{2,3}$-$\pi_\text Y$-$\tau_{2,4}$\big)-$\pi_\text X$-}\nonumber\\
& &
\text{\big($\tau_{3,1}$-$\pi_\text Y$-$\tau_{3,2}$-$\pi_\text Y$-$\tau_{3,3}$-$\pi_\text Y$-$\tau_{3,4}$\big)-$\pi_\text X$-}\nonumber\\
& &
\text{\big($\tau_{4,1}$-$\pi_\text Y$-$\tau_{4,2}$-$\pi_\text Y$-$\tau_{4,3}$-$\pi_\text Y$-$\tau_{4,4}$\big)-$\pi_\text X$}
\end{eqnarray}
where $\tau_{i,j}$ denotes the $j$-th delay in the $i$-th UDD in the whole sequence.
Therefore, QDD(ZY) of odd level is actually based on pulses about $x$ and $y$ axes,
and has a structure similar to the periodic sequences based on $x$ and $y$ rotations,
which are known to be very robust with respect to the pulse errors \cite{deLange10,Ryan10,Tyryshkin10,Ajoy10,Wang10,Gullion90}.
Such a structure can not be achieved with QDD based on $\pi_X$ and $\pi_Y$ pulses.
At even level, the $x,y$-based QDD lacks a final $\pi_X$ pulse, see Eq.~\ref{eq:seq_QDDeven}.
At odd level, the $x,y$-based QDD has even number of $\pi_Y$ pulses sandwiched by $\pi_X$ pulses, see
Eqs.~\ref{eq:seq_QDDodd} and \ref{eq:qdd3example}, while for robustness we need odd
number of $y$-pulses.

In the limit $t\to 0$, to first order in the pulse errors,
the evolution operator for QDD(ZY) of level $\ell=2n-1$ is
\begin{equation}
U^{\tt QDD(ZY)}_{\ell=2n-1}=(-1)^n [1+2 i n(m_x+n_y)\sigma_z].
\end{equation}
Since for our system $m_x=n_y=0$, the DD performance is determined by accumulation of the higher-order errors.

Therefore, for QDD(ZY) of odd level,
the dominant pulse errors are the in-plane components of  rotation axis errors.
For the system under consideration such errors are negligible, and
QDD(ZY) is a good choice for decoherence suppression for all quantum states.

\section{conclusion}
\label{sec:conclusions}
We have studied accumulation of the pulse errors for two aperiodic dynamical decoupling sequences, UDD and QDD,
using the phosphorus electron spins in silicon as an example of experimental implementation.
We show that the decoupling fidelity strongly depends on the initial state.
At long times, we observe saturation of fidelities.
UDD and QDD of even and odd levels are found to perform differently.


UDD protocol is based on a single-axis control, so the decoupling fidelity
is highest for the spin component along the control axis.
QDD preserves the three spin states with closer fidelities.
In particular, QDD of odd level based on $\pi_Z$
and $\pi_Y$ pulses is very robust with respect to the pulse errors,
due to its special structure.
Our results can be useful for experimental implementations of the aperiodic decoupling
protocols, and for deeper understanding of the influence of errors on quantum control
of spin systems.

\acknowledgements
We would like to thank K. Khodjasteh and L. Viola for useful discussions,
and A. M. Tyryshkin and S. A. Lyon for useful discussions and stimulating suggestions.
Work at the Ames Laboratory was supported by the Department of Energy --
Basic Energy Sciences under Contract No. DE-AC02-07CH11358.


\end{document}